\documentclass[review]{elsarticle}

\usepackage{lineno,hyperref}
\modulolinenumbers[5]
\usepackage[latin9]{inputenc}
\usepackage{amsmath}
\usepackage{amssymb}
\usepackage{graphicx}
\usepackage{subcaption}
\usepackage[section]{placeins}  %Force figure in section

\journal{Wave Motion}

\begin{document}

\begin{frontmatter}

\title{Acoustic cloaking: geometric transform, homogenization and a genetic algorithm}

%% Group authors per affiliation:
\author{Lucas Pomot$^{1,3}$} 
\author{C\'edric Payan \fnref{LMA}} 
\author{Marcel Remillieux \fnref{LANL}}
\author{S\'ebastien Guenneau \fnref{Fresnel}}
%\address{Radarweg 29, Amsterdam}
\fntext[LMA]{Aix Marseille Univ, CNRS, Centrale Marseille, LMA, Marseille, France}
\fntext[LANL]{Geophysics Group (EES-17), Los Alamos National Laboratory, Los Alamos, New Mexico 87545, USA}
\fntext[Fresnel]{Aix Marseille Univ, CNRS, Centrale Marseille, Institut Fresnel, Marseille, France}

%\author[mysecondaryaddress]{Global Customer Service\corref{mycorrespondingauthor}}
%\cortext[mycorrespondingauthor]{Corresponding author}
%\ead{support@elsevier.com}

%\address[mymainaddress]{1600 John F Kennedy Boulevard, Philadelphia}
%\address[mysecondaryaddress]{360 Park Avenue South, New York}

\begin{abstract}
A general process is proposed to experimentally design anisotropic inhomogeneous metamaterials obtained through a change of coordinate in the Helmholtz equation. The method is applied to the case of a cylindrical transformation that allows to perform cloaking. To approximate such complex metamaterials we apply results of the theory of homogenization and combine them with a genetic algorithm. To illustrate the power of our approach, we design three types of cloaks composed of isotropic concentric layers structured with three types of perforations: curved rectangles, split rings and crosses.  These cloaks have parameters compatible with existing technology and they mimic the behavior of the transformed material. Numerical simulations have been performed to qualitatively and quantitatively study the cloaking efficiency of these metamaterials.
\end{abstract}


\end{frontmatter}

%\linenumbers

\section{Introduction}
The geometrical transformation of the Maxwell equations has been studied theoretically back in 1962 by Post \cite{Post1962} and numerically since 1994 by Nicolet \textit{et al.} \cite{Nicolet1994} and 1996 by Ward and Pendry \cite{Ward1996}. Those studies were first focused on the electromagnetic equations. More recent results by Greenleaf \textit{et al.} \cite{Greenleaf2003} (2003), Leonhardt \cite{Leonhardt2006} (2006) and Pendry \textit{et al.} \cite{Pendry2006} (2006) showed that these transformations can be used to perform cloaking on electromagnetic waves. Those results of great interest have then been used when studying acoustic waves which satisfy a similar equation of propagation to the electromagnetic waves. In this case, geometrical transformation introduced either an anisotropic mass density (Cummer \textit{et al.} \cite{Cummer2007} 2007, Chen \textit{et al.} \cite{Chen2007} 2007, Cummer \textit{et al.} \cite{Cummer2008} 2008) or an anisotropic inertia (Milton \textit{et al.} \cite{Milton2007} 2007). Such medium property can arise from microstructures as discussed in Mei \textit{et al.} \cite{Mei2007} (2007) and Torrent \textit{et al.} \cite{Torrent2008} (2008). A challenge for experimental application is then to find the appropriate microstructure that properly mimics the transformed parameters. In this paper we propose a process to efficiently determine this microstructure.\\
After recalling the elementary results of geometrical transformation and homogenization, we demonstrate how a genetic algorithm can be used to design the microstructure associated with a given geometrical transformation. This process will be illustrated using a cylindrical transformation that 'creates a hole in space' by mapping a disc on an annulus and two types of designs will be determined, each of them composed of an elementary cell and a variation of mass density and bulk modulus. Looking at those variations we show that certain microstructures seem to be more convenient for experimental application as they require a smaller range of elastic parameters.
\section{Elementary results concerning geometrical transformation}
\begin{figure*}
\includegraphics[width=\textwidth]{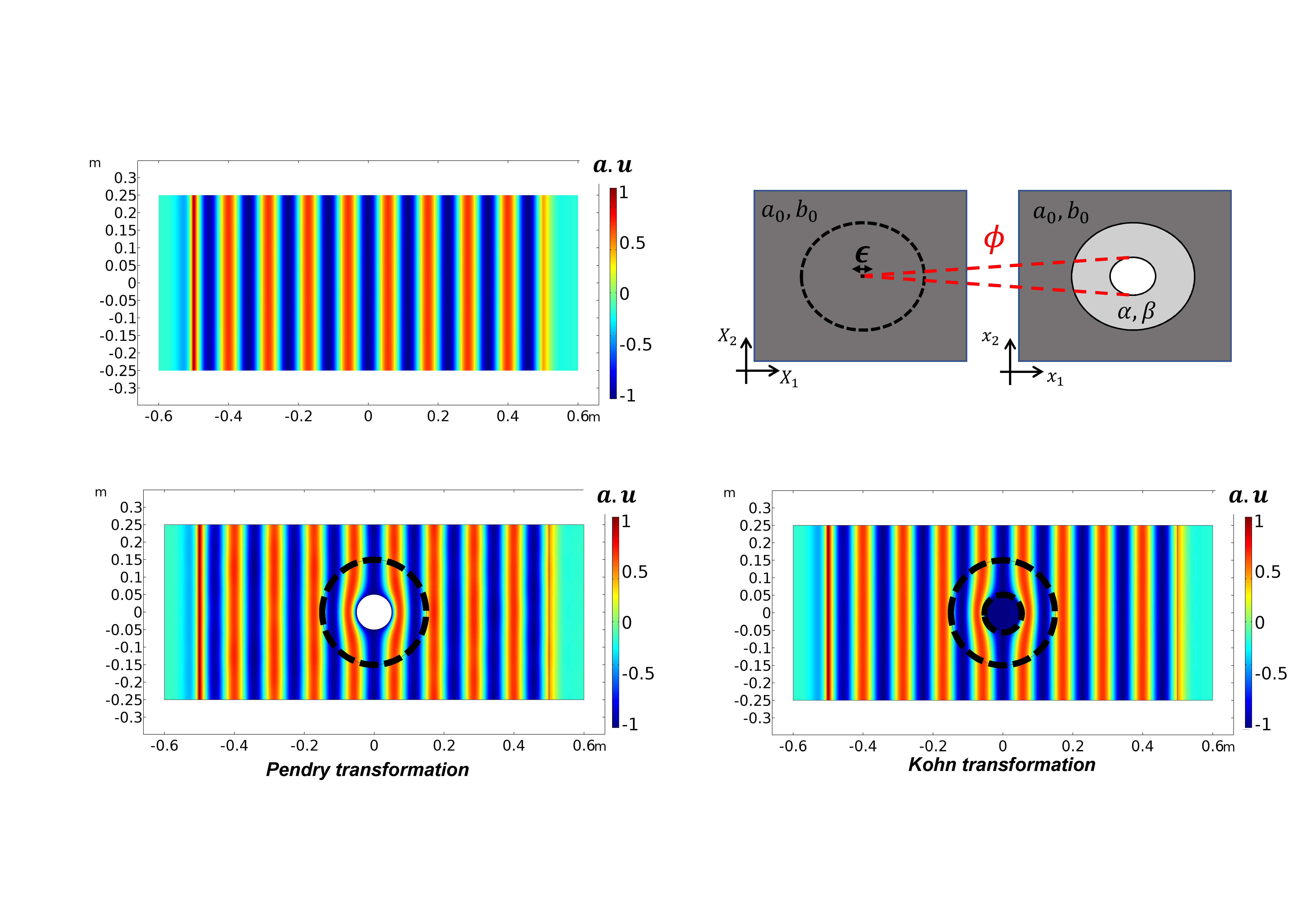}
\caption{Upper Left) Propagation of a monochromatic wave from left to right in an homogeneous medium defined by $a = 1$ and $b = 4.45 \ 10^{-7}$ with PML at $x = 0.5m$ and $x = -0.5m$. Upper Right) Schematic representation of the geometrical transformation applied to the homogeneous medium. The transformation sends a disk of radius $\epsilon$ to a disk of radius $R_{int}$. Lower Left) and Lower Right) Numerical results obtained for both a Kohn's (Lower Left) and a Pendry's (Lower Right) transformation. The form invariant property of the equation of propagation is clearly visible as the wavefield is distorted only in the region affected by the transformation. Note the presence of perfectly matched layers on the left and right hand sides. }  
 \label{fig1}
\end{figure*}

In this section, we recall some elementary results on geometrical transformation in the case of a time-harmonic wave equation of the form: 
\begin{equation}
\nabla\cdot a(\textbf{X}) \nabla u(\textbf{X}) + b(\textbf{X}) \omega^2 u(\textbf{X}) = 0
\label{eq:prop}
\end{equation}
with $a$ and $b$ the spatially varying parameters that describe the medium of propagation.
This so-called Helmholtz equation appears in various fields of physics such as acoustic propagation or electromagnetism. For instance, if we consider anti-plane shear waves propagating within a a solid elastic medium invariant along the anti-plane direction, then $a$ stands for the shear modulus and $b$ stands for the mass density of this medium \cite{Guenneau2010}; likewise if we consider transverse magnetic waves in a dielectric medium invariant along the anti-plane direction, $a$ stands for the inverse of electric permittivity and $b$ stands for the magnetic permeability \cite{Farhat2008}; if we consider water waves, $a$ can stand for water depth, in which case $b=1$ \cite{Maurel2013}, or $a$ stands for the product of the phase and the group density and $b$ for the ratio of the group velocity over the phase velocity in the context of the mild-slope equation \cite{Dupont2016}. In the present paper, we choose to study pressure acoustic waves propagating within a non-viscous fluid, and so $a$ stands for the inverse of mass density and $b$ for the inverse of bulk modulus but our results can be easily translated to the fore mentioned wave areas.
We now apply a coordinate transformation of the form $\phi : \textbf{X} \rightarrow \textbf{x}$ on the domain of propagation. Using the results in \cite{Norris2008} it is a straight forward matter to obtain the governing equation: 
\begin{equation}
\nabla\cdot \alpha \nabla u(\textbf{x}) + \beta \omega^2 u(\textbf{x}) = 0
\label{eq:prop2}
\end{equation}
where $\alpha = a J J^T \text{det} J$ is now a matrix valued spatially varying parameter that depends on the Jacobian $J_{ij} = \partial x_i / \partial X_j$ of the geometrical transformation $\phi$ and $\beta = b \ \text{det} J$ is a spatially varying scalar parameter. The main observation here is that the equation of propagation is form invariant meaning that the transformed equation still describes the same physical phenomena but not in the same medium of propagation. Let us focus our analysis on the following non-linear geometrical transformation which is written in polar coordinates: 
\begin{equation}
\phi : \textbf{X} \rightarrow (r, \theta) \rightarrow (r' = \sqrt{ \frac{R_2^2 - R_1^2}{R_2^2} r^2 + R_1^2 }, \ \ \theta' = \theta) \rightarrow \textbf{x}.
\label{eq:TG}
\end{equation}
This geometrical transformation introduces an annulus of anisotropic inhomogeneous material of internal radius $R_1$ and external radius $R_2$ (see figure \ref{fig1} (Upper Right)) in the propagation domain. In theory this transformation sends a point to a disk of radius $R_1$, however this approach introduces a divergence in the properties of the transformed medium. In practice it is preferable to send a disk of radius $\epsilon$ very small in comparison with the other characteristic lengths onto a disk of radius $R_1$. If the initial disk of size $\epsilon$ is composed of the same material as the surrounding medium we call this transformation Kohn's transformation. If the initial disk is empty, meaning it is a perforation in the medium of propagation with homogeneous Neumann boundary condition (i.e. zero normal derivative that is zero flux at the boundary), we call the transformation Pendry's transformation. Let us note that such Neumann boundary conditions hold in our acoustic case for rigid cylinders \cite{Torrent2006}, as well as in the water wave case \cite{Dupont2016}. However, this model holds for void cylindrical inclusions in the case of anti-plane shear waves, and infinite conducting cylinders in the case of transverse magnetic waves. From now on, we denote by perforations such Neumann types inclusions.
It is useful to consider three compound geometric transforms: first mapping Cartesian coordinates onto polar coordinates with Jacobian $J_{xr}=\frac{\partial(r,\theta)}{\partial(x_1,x_2)}$, then performing the transformation introduced in equation \ref{eq:TG} with Jacobian $J_{rr'}=\frac{\partial(r',\theta')}{\partial(r,\theta)}$ and finally mapping back from the stretched polar coordinates to the stretched Cartesian coordinates with Jacobian $J_{r'x'}=\frac{\partial(x', y')}{\partial(r',\theta')}$. The derivation of the total Jacobian $J_{xx'}$ and its determinant are given by:
\begin{equation}
\begin{split}
J_{xx'} &= J_{xr} J_{rr'} J_{r'x'} \\
&= R(\theta) \text{diag(1,r)} J_{rr'} \text{diag(1,1/r')} R(\theta') \\
&= R(\theta) \begin{bmatrix} \frac{r'}{\frac{R_2^2 - R_1^2}{R_2^2} r} & 0 \\ 0 & \frac{r}{r'} \end{bmatrix} R(\theta')\\
\text{det}J &= \frac{R_2^2}{R_2^2 - R_1^2}
\end{split}
\end{equation}
$R(\theta)$ being the rotation matrix through an angle $\theta$.
The main advantage of this non linear geometrical transformation in comparison with Pendry's or Kohn's transformation is that the determinant of the Jacobian is constant thus the transformed medium parameter $\beta = b \ \text{det}J$ is constant. The new medium parameters are
\begin{equation}
\begin{split}
\alpha &= a J^{-1} J^{-T} \text{det}J = a R(\theta) \begin{bmatrix} \frac{r'^2 - R_1^2}{r'^2} & 0 \\ 0 & \frac{r'^2}{r'^2 - R_1^2} \end{bmatrix} R(\theta')\\ 
\beta &= b \ \text{det}J = b \frac{R_2^2}{R_2^2 - R_1^2}\\
\label{eq:}
\end{split}
\end{equation}
where $J^T$ denotes the transpose of $J$. We implemented these anisotropic inhomogeneous medium parameters in the COMSOL numerical simulation software. The results are shown on figure \ref{fig1}(Lower Left), and figure \ref{fig1}(Lower Right). The control of wave propagation achieved here could have several applications in various fields of physics. However, considering the characteristics of the transformed material, such medium is really hard to create in practice. What we suggest instead is to structure the medium of propagation on a microscopic level (according to the field of physics that we consider) and to use classical results of the theory of homogenization to determine and tune the effective properties of the medium. 

		\FloatBarrier
\section{Inverse homogenization using a genetic algorithm}
\subsection{Homogenization of the Helmholtz equation \label{sec:Homogenization}}
In order to mimic these anisotropic inhomogeneous materials mathematically introduced in the previous section we apply classical results of the theory of homogenization in order to determine the effective parameters of a complex medium with small perturbations. The aim of this approach is to perform what we wish to call an inverse homogenization (or retrieval method \cite{Smith2005}, \cite{Cherkaev2001}, \cite{Maurel2013}, \cite{Fokin2007}), where we tune the characteristics of the small perforations to obtain an effective medium that corresponds to the medium given by the change of variable. Homogenization results used here were derived using asymptotic expansions as in \cite{Bensoussan1978} and thus only the main results are recalled here for two different cases. 
\subsubsection*{Homogenization of a 1D laminar lattice}
\begin{figure*}
        \centering
        \begin{subfigure}[b]{0.55\textwidth}
            \centering
            \includegraphics[width=\textwidth]{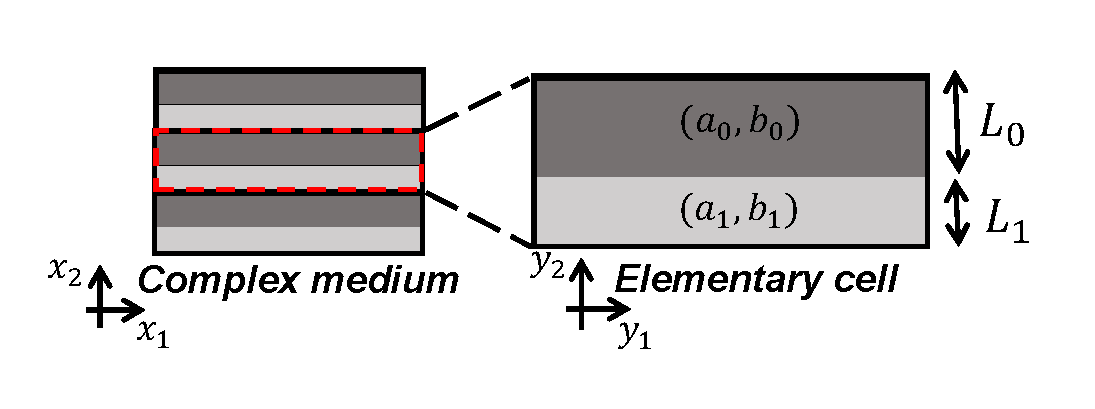}
            \caption{}  
            \label{fig2a}
        \end{subfigure}
        \hfill
        \begin{subfigure}[b]{0.55\textwidth}  
            \centering 
            \includegraphics[width=\textwidth]{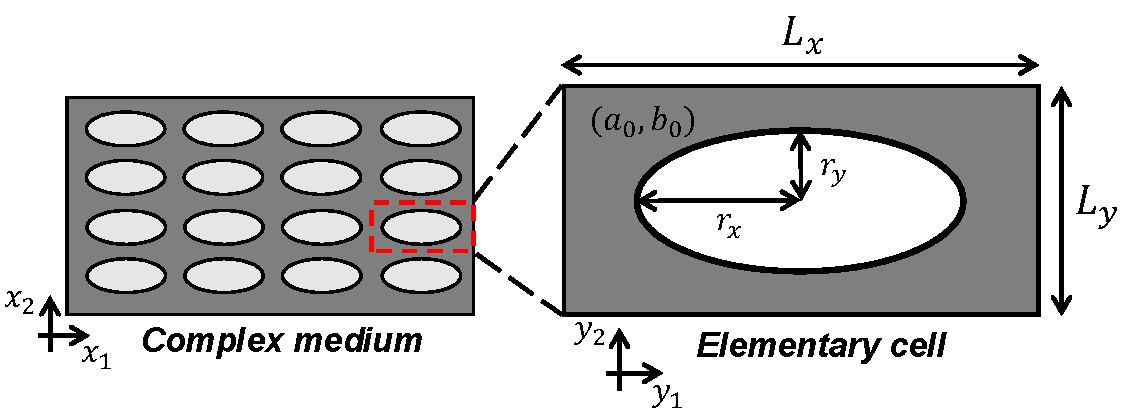}
            \caption{}   
            \label{fig2b}
        \end{subfigure}
				\caption{a) Schematic representation of the 1D laminar lattice define by 6 parameters if $L_0 \neq L_1$ and 4 parameters if $L_0 = L_1$ . b) Schematic representation of the 2D periodic rectangular lattice define by 6 parameters. We use those two lattices to mimic the desired medium through homogenization.}  
        \label{fig2}
    \end{figure*}

\begin{figure*}
        \centering
        \begin{subfigure}[b]{0.45\textwidth}
            \centering
            \includegraphics[width=\textwidth]{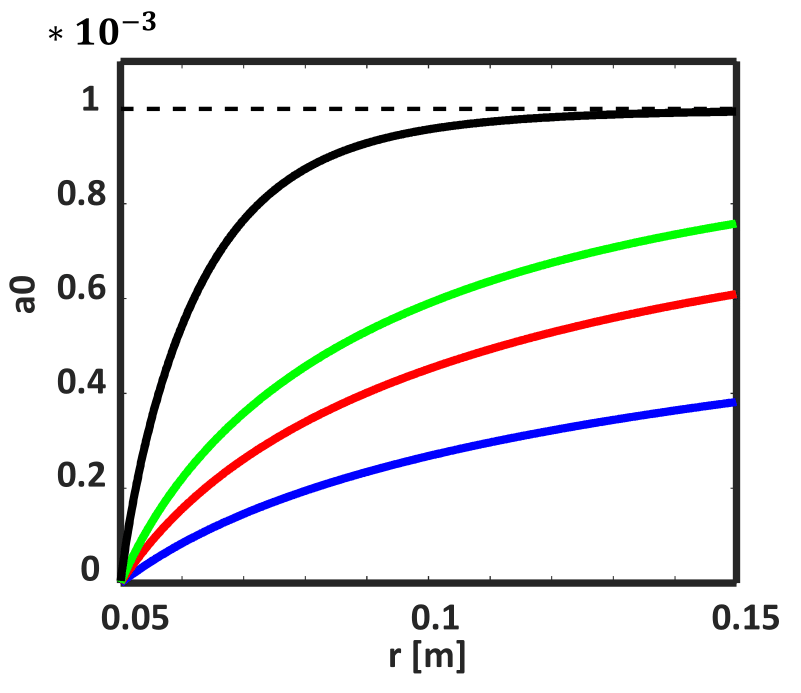}
            \caption{}  
            \label{fig:a0ForVariousTG}
        \end{subfigure}
        \begin{subfigure}[b]{0.45\textwidth}  
            \centering 
            \includegraphics[width=\textwidth]{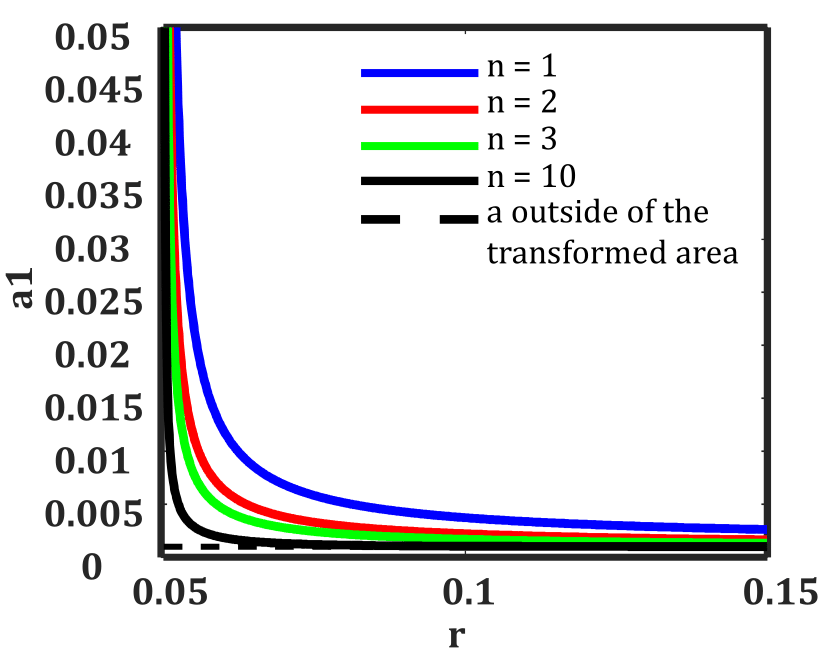}
            \caption{}   
            \label{fig:a1ForVariousTG}
        \end{subfigure}
\caption{Variation of homogenized parameters $a_0$ (a) and $a_1$ (b) depending on the order $n$ of the transformation. Note that the higher the order of the transformation the faster the homogenized parameters $(a_0,a_1)$ get close to the parameter $a$. However also the larger the variation for $r$ close to $R_{int}$. Regarding experimental applications, the high order transformations should be considered when it is doable to create an intense and localized change in the properties of the medium propagation. On the other hand, if it is easier to create a slow gradient in the propagation medium properties one should consider a low order transformation.}  
\label{fig:a0&a1ForVariousTG}
\end{figure*}

A simple example is the case of an alternation of two isotropic propagation media (see figure \ref{fig2a}). Analytical formulae can then be deduced to determine the effective properties of the effective anisotropic medium \cite{Bensoussan1978}:
\begin{equation}
\begin{split}
\alpha &= \begin{bmatrix} \langle a^{-1} \rangle ^{-1}  & 0 \\ 0 & \langle a \rangle \end{bmatrix} \\
\beta &= \langle b \rangle
\label{eq:param_hom1D}
\end{split}
\end{equation}
where $ \langle a \rangle = L_0 a_0 + L_1 a_1$, $L_0$ and $L_1$ being the thickness of the two layers that compose the microstructure (see figure \ref{fig2a}). In this configuration the tuning parameters are the different homogeneous isotropic media defined by $(a_i,b_i)$. Possible additional tuning parameters are the size $L_0$ and $L_1$ of each layer. In the following numerical results we choose to set $L_0 = L_1$. To derive the numerical value of $a_1$ and $a_2$ we solve the following system \cite{Petiteau2014}:
\begin{equation}
\begin{cases}
\alpha_{rr} = \langle a^{-1} \rangle ^{-1} = \frac{a_0 a_1}{a_1 L_0 + a_0 L_1} \\
\alpha_{\theta \theta } = \langle a \rangle = L_0 a_0 + L_1 a_1 \\
\beta = \langle b \rangle = L_0 b_0 + L_1 b_1
\end{cases}
\end{equation}
In the case $L_0 = L_1 = 0.5$ the expression of $\alpha_{\theta \theta }$ and $\alpha_{rr}$ reduced to:
\begin{equation}
\begin{cases}
 a_0 = \alpha_{\theta \theta } - \sqrt{\alpha_{\theta \theta } ^2 -\alpha_{\theta \theta }\alpha_{rr} } \\
 a_1 =  \alpha_{\theta \theta } + \sqrt{\alpha_{\theta \theta } ^2 -\alpha_{\theta \theta }\alpha_{rr} }\\
\end{cases}
%\end{equation}
\label{eq:homogeneizationSystem}
\end{equation}
		\FloatBarrier
We note here that we can solve this system in the $(r, \theta)$ domain if we then map the obtained microstructure to the $(x,y)$ domain using a conformal map, a point we will develop in section (4.1). The only terms in this system of equations that are dependent on the geometrical transformation are the left hand terms. Previously we justified the choice of our transformation (\ref{eq:TG}) by arguing that the determinant of the associated Jacobian is constant and thus we have a constant $\beta$. However we did not take into account the influence of this transformation on the values of the homogenized elastic parameters $a_0$, $a_1$, $b_0$, $b_1$. We discuss here the values taken by those parameters for different transformations defined by: 
\begin{equation}
\phi^n : (r, \theta) \rightarrow (r' = \Big(\frac{R_2^n - R_1^n}{R_2^n} r^n + R_1^n \Big)^{\frac{1}{n}}, \theta' = \theta) 
\label{eq:TGPowerN}
\end{equation}
where $n$ is a positive integer. $n = 1$ is the classic linear Pendry's transformation, $ n =2$ is the quadratic transformation that we consider in this paper. We then compute the parameters $a_0(r')$ and $a_1(r')$ for $n = 1,2,3,10$. The results are shown in figure \ref{fig:a0&a1ForVariousTG}. To choose the parameters $b_0$ and $b_1$ we have an under-determined system, meaning that we have an extra degree of freedom to pick those. The simplest solution would be to take $b_0 = b_1 = \beta$ but depending on experimental realization it may not be the best choice as we will see later.
\FloatBarrier
\subsubsection*{Homogenization of a 2D periodic rectangular lattice}
We consider here a complex medium made of identical elementary cells regularly spaced, see figure \ref{fig2b}. To model this type of medium we introduce a microscopic scale described by the variable $\textbf{y}$ in addition to the macroscopic scale described by $\textbf{x}$. The effective parameter of such a medium can be numerically calculated. We give here the analytical formulae of these effective parameters \cite{Bensoussan1978}, \cite{Milton2002}:
\begin{equation}
\begin{split}
\alpha^H &= \begin{bmatrix} \langle a \rangle - \langle a \partial_{y_1}w_1 \rangle & -\langle a \partial_{y_1} w_2 \rangle \\ - \langle \partial_{y_2} w_1 \rangle & \langle a \rangle - \langle a \partial_{y_2}w_2 \rangle \end{bmatrix}\\
\beta^H &= \langle b \rangle
\end{split}
\label{eq:param_hom}
\end{equation} 
where $\langle \cdot \rangle$ is the mean operator over an elementary cell, $\partial_{y_1}w_1$ design the partial derivative if $w_1$ by $y_1$ and $(w_1,w_2)$ are the solutions of the annex problem solved on only one elementary cell with periodic boundary conditions
\begin{equation}
\nabla_{{\bf y}} \cdot \big( a (\nabla_{{\bf y}} w_j ({\bf y}) + {\bf e_j}) \big) = 0
\end{equation}
with ${\bf e}_1=(1,0)$ and ${\bf e}_2=(0,1)$ the vectors of the basis. In this case the effective properties of the medium can be tuned using several parameters such as the form of the perforations (elliptic in our case), the medium properties or the size of the elementary cell. All these parameters are described in figure \ref{fig2b}. The objective is now to find the correct set of parameters that makes it possible to properly mimic the medium of propagation. We called this process inverse homogenization: we tune the elementary cell to obtain the desired homogenized medium of propagation. Considering the number of tuning parameters and the lack of completely analytical formula we implement a genetic algorithm that is well suited to solve such inverse problems. 
\\
\subsection{Genetic algorithm}
The inverse problem is solved using a genetic algorithm (GA) proposed in \cite{Haupt2004}. The main advantage of GA over other optimization process is its efficiency to find a global minimum within a large and discrete solution space. In our case we try to minimize the following cost function: 
\begin{equation}
\gamma (\alpha^H) = \max_{\{ij\}} \frac{|\alpha_{ij} - \alpha_{ij}^H|}{|\alpha_{ij}|} 
\label{eq:costfun}
\end{equation}
for various sets of parameters (the parameters we considered are defined on figure \ref{fig2}). In this formula $\alpha$ defines the characteristics of the medium we want to approximate and $\alpha^H$ the homogenized medium for a particular set of parameters. The GA will be used to perform stochastic search based on the principles of natural selection and evolution. An initial population $N$ of individuals is going to be generated, each individual being defined by its set of parameters. For the initial iteration the parameters are chosen randomly. The homogenization is then performed for each set of parameters (meaning that we solve the annex problem for each set of parameters) and evaluated using the cost function $\gamma$ defined in equation (\ref{eq:costfun}). The closer the homogenized medium to the desired medium, the smaller the cost function. The individuals who did poorly at this iteration are then eliminated and new sets of parameters are determined based on the individuals that survived the previous iteration. Each parameter can be compared to a gene and the set of parameters to a DNA (which takes its name from biology). To determine the DNA of the next generation we take the arithmetic means of the parents's genes, meaning that if parent $i$ is defined as $p^i = \{a^i, b^i, L_x^i, L_y^j, r_x^i, r_y^i \} = DNA(p^i)$ and parent $i +1 $ as $p^{i+1} = DNA(p^{i+1})$, then the genes of children $c^i$ are defined as $DNA(c^i) = \mu DNA(p^i) + (1-\mu) DNA(p^{i+1})$ where $\mu$ is an arbitrary real number between 0 and 1.  To avoid being trapped in local minima we add a mutation factor who takes a chosen percentage of genes picked randomly and attribute them random values. Thus, if the algorithm converges to a minimum that is not a global minimum a random mutation can still create an individual with a better performing DNA. This individual can then influence the next generations. There can be a lot of tweaking in GA to choose the ideal population number, mutation rate or other parameters that we did not introduce here for sake of simplicity. However the GA performed really well in our case and allowed us to determine quickly a large number of elementary cells that properly mimic our transformed medium after approximately 15 iterations. 
\\
\section{Illustrative numerical results}
\subsection{Conformal mapping }
\begin{figure}
\centering
\includegraphics[width=0.9\columnwidth]{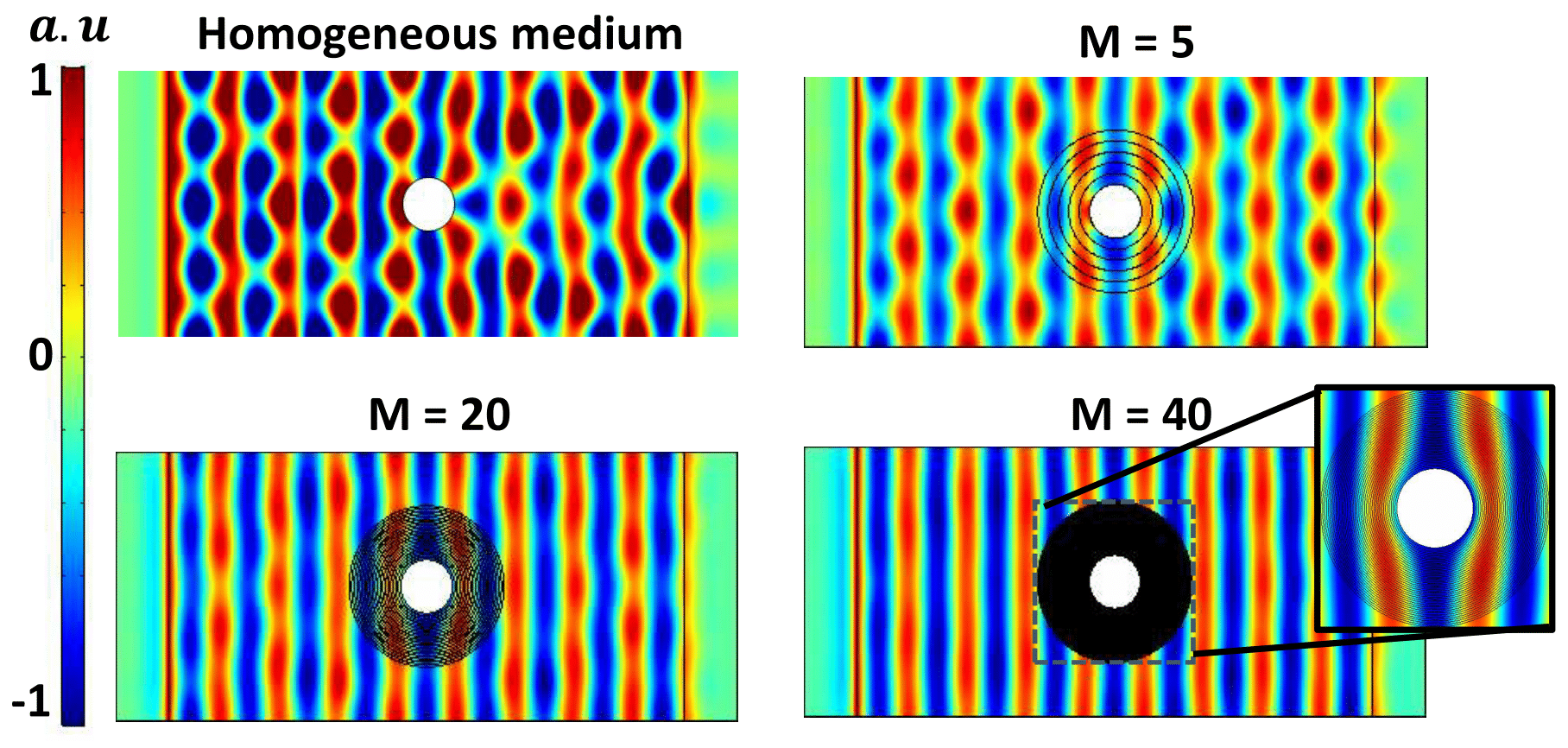}
\caption{Qualitative results for various choices of subdivision M of the anisotropic inhomogeneous medium into anisotropic homogeneous medium. Note the visually good results obtained from M = 20 but note also that M = 5 gives already a noticeable reduction of scattering in comparison with the obstacle in the homogeneous medium. }
\label{figChoiceOfM}
\end{figure}

In section \ref{sec:Homogenization} we quickly introduced the fact that the microstructures are designed in the $(r, \theta)$ coordinates whereas we want to obtain a final microstructure in the $(x,y)$ coordinates. In fact the inverse homogenization will take place in the $(r,\theta)$ coordinates before being mapped on the $(x,y)$ coordinates using a conformal map, a geometrical transformation that does not impact the medium properties. The conformal map we consider maps a rectangular domain $[R_1, R_2] \times [-\pi/\eta, \pi/\eta]$ onto an annular domain $[R_1, R_2] \times [0, 2 \pi]$ \cite{Dupont2016}: 
\begin{equation}
w = \psi e^{\eta (x + iy)}, \ \ \text{with} \ \ \ \eta = \frac{ln(R_2/R_1)}{R_2 - R_1} \ \ \ \psi = \frac{R_2 + R_1}{e^{\eta R_2} + e^{\eta R_1}}
\label{eq:TC}
\end{equation}
It is important to note here that the medium given by the change of variable is inhomogeneous in addition to being anisotropic whereas homogenization theory can only achieve homogeneous anisotropic media. To achieve the required inhomogeneity we approximate the inhomogeneous anisotropic medium by several homogeneous anisotropic media. The impact of this approximation was quantitatively studied in figure \ref{figChoiceOfM}. Based on these results we choose to divide our medium into M = 20 homogeneous anisotropic media. However it is important to notice that this approximation does already a fair job for $M = 5$. A similar question arises when choosing the number $N$ of elementary cells needed to properly describe an homogeneous anisotropic material. In our case we found that N = 3 elementary cells for the laminar case and only N = 1 elementary cell for the periodic rectangular lattice give quantitatively good results.\\
\FloatBarrier
\subsection{1D laminar case}
\begin{figure*}
        \centering
        \begin{subfigure}[b]{0.47\textwidth}
            \centering
            \includegraphics[width=\textwidth]{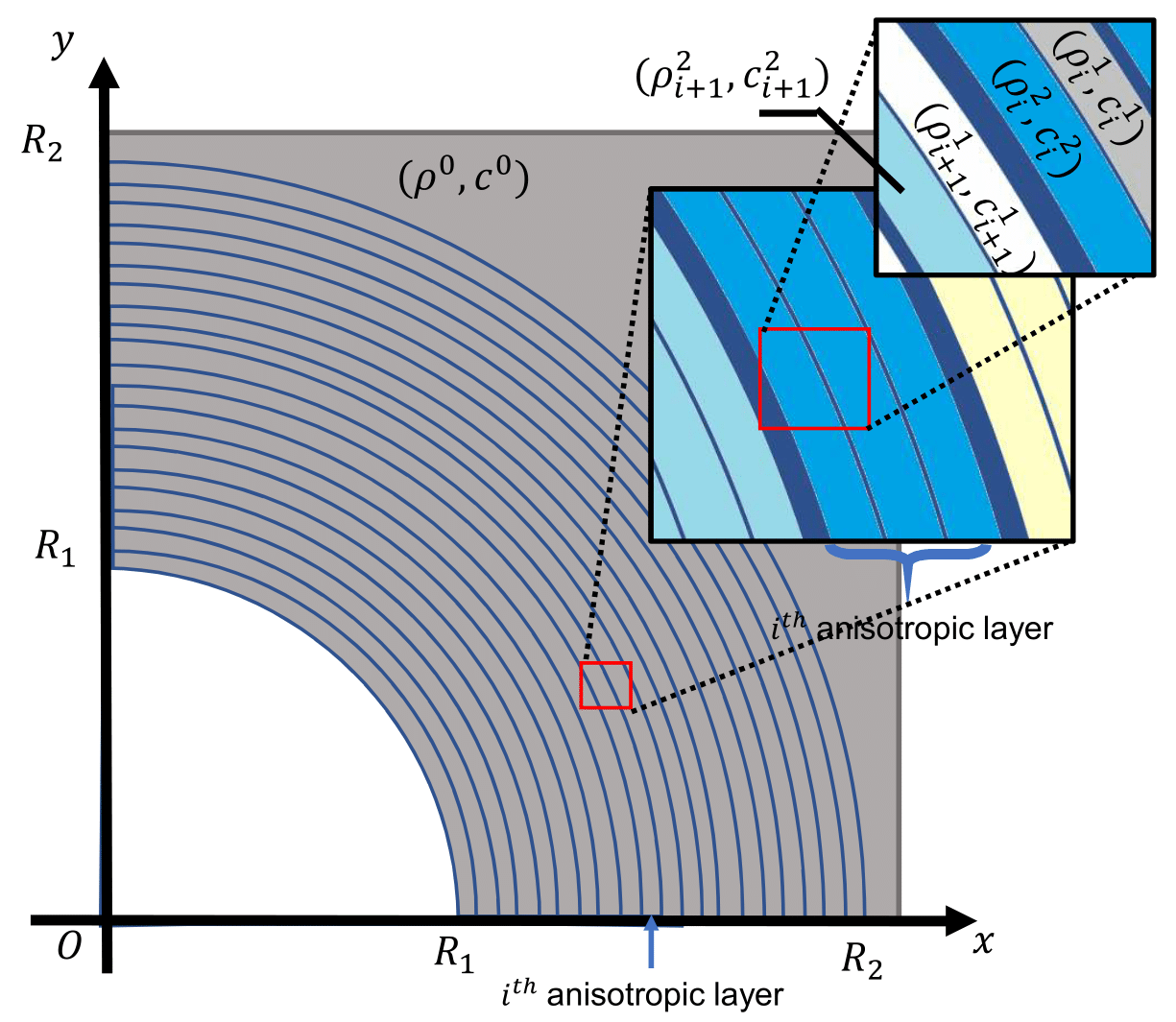}
            \caption{}  
            \label{fig3a}
        \end{subfigure}
        \begin{subfigure}[b]{0.47\textwidth}  
            \centering 
            \includegraphics[width=\textwidth]{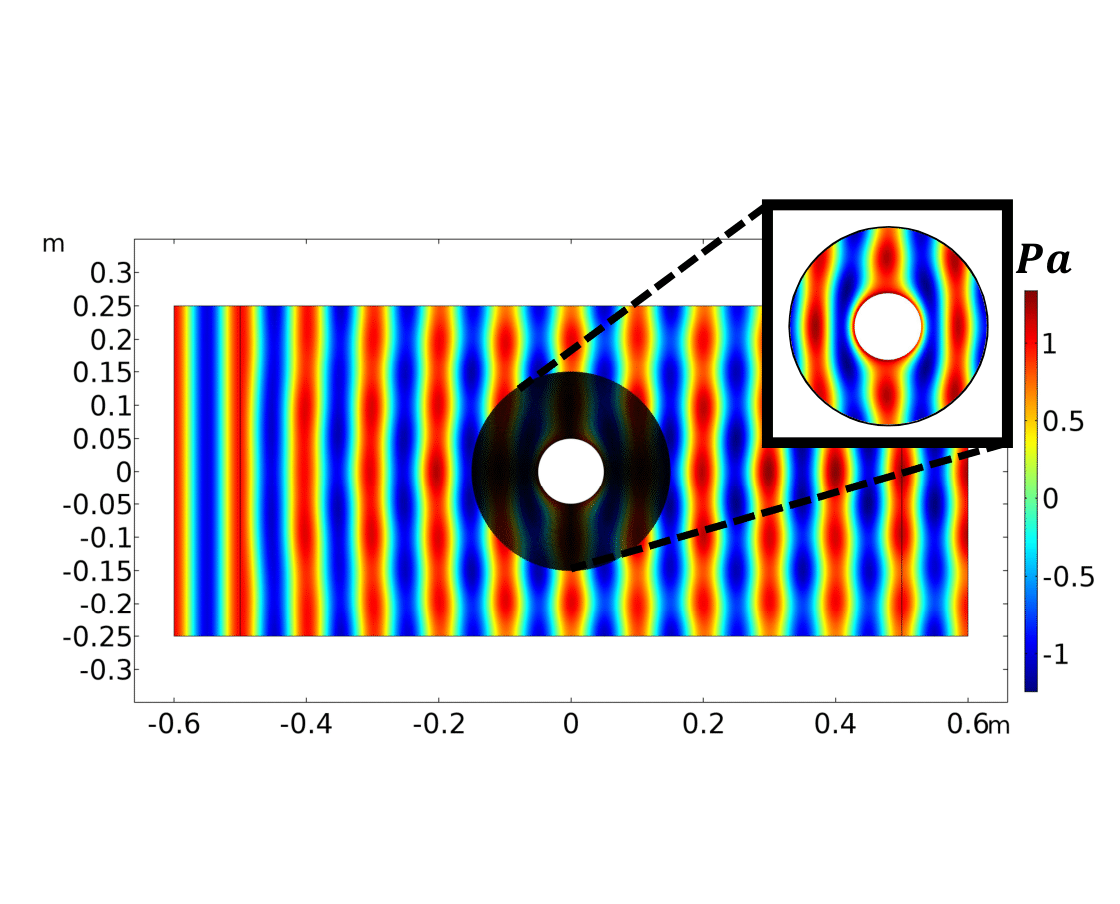}
            \caption{ }   
            \label{fig3b}
        \end{subfigure}
				\caption{a) Representation of the 1D laminar lattice made of M = 20 anisotropic homogeneous media, each of them consisting of 3 alternations of homogeneous isotropic media. In total,  the cloak is made of 120 alternations of homogeneous isotropic rings made of 40 different media. b) Numerical result: Propagation of a monochromatic acoustic wave of frequency $f = 15$ kHz through the homogenized cloak designed with the 1D laminar lattice, surrounded by water. We observe that the process of inverse homogenization we developed allowed us to mimic properly the inhomogeneous anisotropic medium introduce through the geometrical transformation, using only homogeneous isotropic media.}  
        \label{fig3}
\end{figure*}

The calculation was performed using the acoustic module of Comsol which solves the following equation: 
\begin{equation}
\nabla \cdot (\frac{1}{\rho} \nabla p) + \frac{1}{\rho c^2} \omega^2 p = 0,
\label{eq:prop_acous}
\end{equation}
with $p$ the acoustic pressure, $\rho$ the mass density and $c$ the velocity. In fact, one could have alternatively introduced the bulk modulus $B = \rho c^2$ to describe the medium. However since we address here the wave community we find it more natural to consider medium density and wave velocity as parameters. Upon inspection of equation \ref{eq:prop} we have:
\begin{equation}
\begin{split}
\rho &= a^{-1} \\
c &= \sqrt{\frac{a}{b}}
\end{split}
\end{equation}
We can now use the entire method described above to design an invisibility cloak for acoustic wave propagating in water ($\rho_0 = 1000 kg/m^3$,$c_0 = 1500 m/s$) for example. To show the generality of the method we assume that $a$ and $b$ can take any value. It is of course not the case in practice (we do not have a full control on the velocity spatially for example) but depending on the tuning parameters approximations are possible. One of this compromise that we quickly introduce before is the fact that we have an extra degree of freedom on the choice of the parameters $(b_0, b_1)$ which are not entirely define by the system \ref{eq:homogeneizationSystem}. Thus we can use this extra degree of freedom to tune the velocity in order to approach more realistic materials. We consider the geometrical transformation (\ref{eq:TG}) with $R_1 = 0.05$ and $R_2 = 0.15$. The anisotropic inhomogeneous medium is first divided into 20 anisotropic homogeneous media. The properties of each of these anisotropic media are mimicked by using two homogeneous isotropic media and the relation (\ref{eq:param_hom1D}). The elementary cells, defined by the alternation of two homogeneous isotropic media, are repeated 3 times in each anisotropic homogeneous concentric layer along the radial direction. In summary the cloak is made of 40 different media, each of them repeated three times for a n overall structure consisting of an alternation of 120 homogeneous isotropic rings. The complete design of the cloak is shown on figure \ref{fig3a} and the values of the media parameters are given in table \ref{table1}. We stress that the extreme values of the elastic parameters are needed here for a cloak of very high efficiency. As a comparison we show in table \ref{table1forM5} the values needed for a cloak made of $M=5$ layers, which are more realistic in terms of existing media. The numerical result is given in figure \ref{fig3b} for a frequency of 15kHz even if we would like to stress that this design is available for any frequency as long as the size of an elementary cell is small in comparison with the wave wavelength.\\
\FloatBarrier
\subsection{2D rectangular lattice}
\begin{figure*}
        \centering
        \begin{subfigure}[b]{0.4\textwidth}
            \centering
            \includegraphics[width=\textwidth]{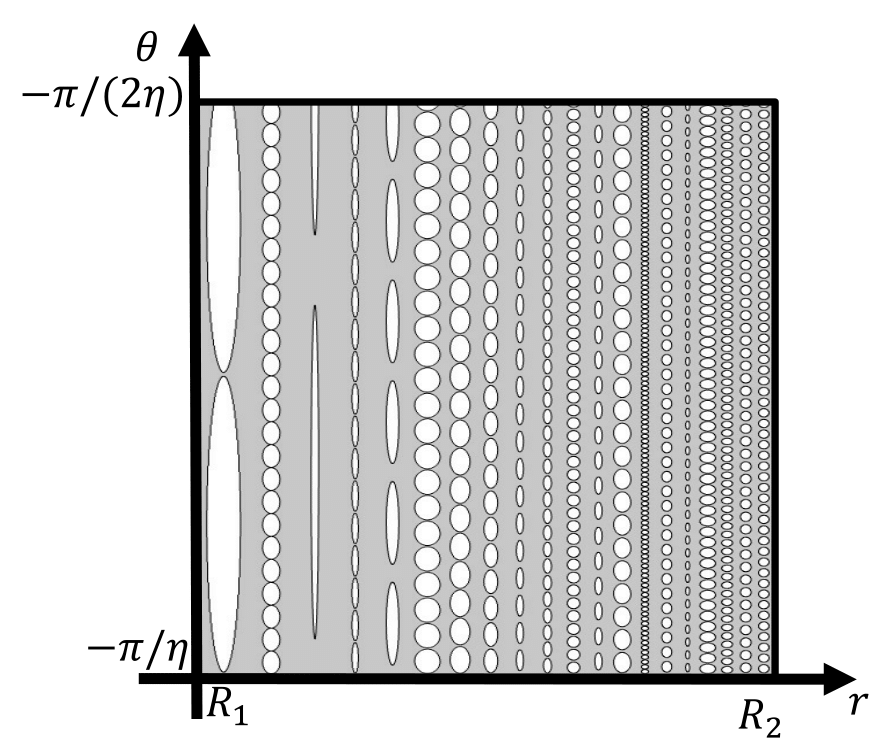}
            \caption{}  
            \label{fig4a}
        \end{subfigure}
        \begin{subfigure}[b]{0.4\textwidth}  
            \centering 
            \includegraphics[width=\textwidth]{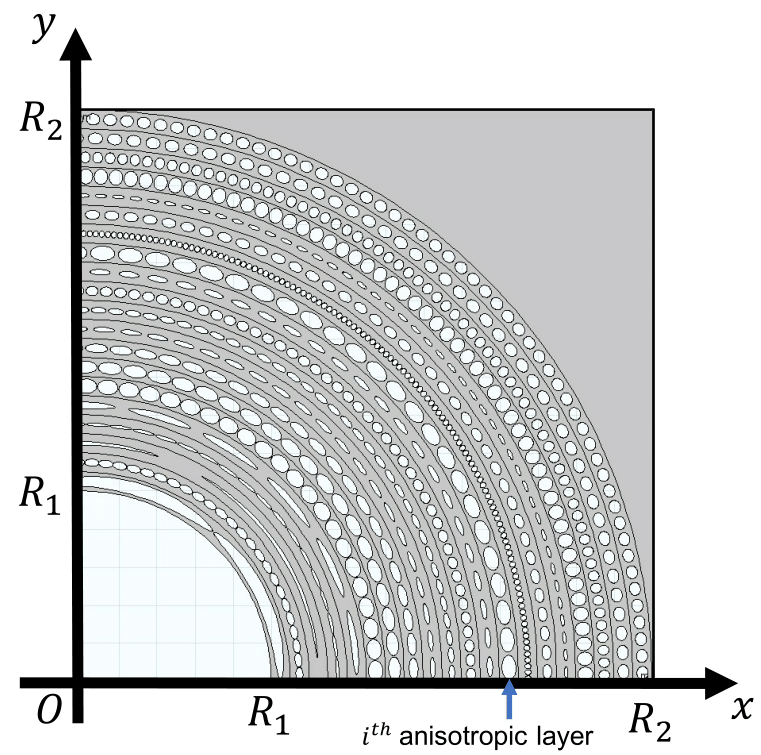}
            \caption{}   
            \label{fig4b}
				\end{subfigure}
%				\baselineskip
				\begin{subfigure}[b]{0.8\textwidth}  
            \centering 
            \includegraphics[width=\textwidth]{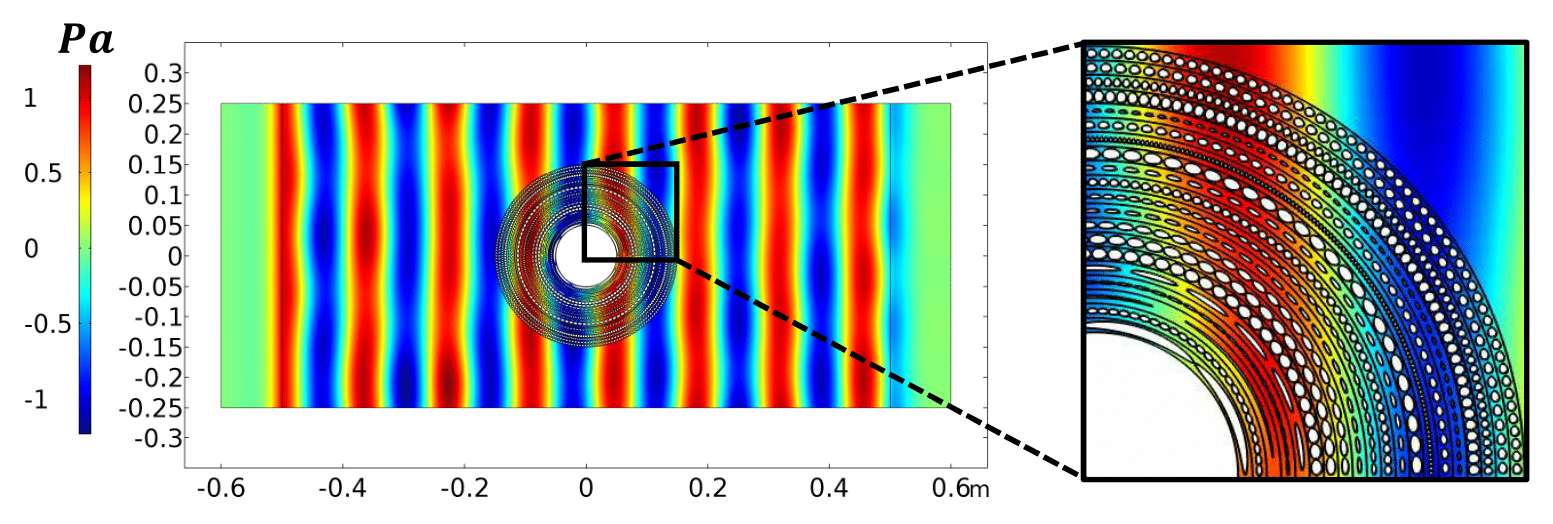}
            \caption{}   
            \label{fig4c}
        \end{subfigure}
				\caption{a) Design of the microstructure of the cloak in the $(r, \theta)$ domain, before the application of the conformal map defined by equation (\ref{eq:TC}). b) Design of the microstructure of the cloak in the $(x, y)$ domain, after the conformal mapping. c) Numerical result: propagation of a monochromatic acoustic wave of frequency $f = 13$ kHz through the homogenized cloak designed with the 2D rectangular lattice, and surrounded by water. Once again we mimic properly the inhomogeneous anisotropic medium introduced through the geometrical transformation.} 
        \label{fig4}
\end{figure*}

In a similar way we can work out a design using a different microstructure. In the following we still consider the propagation of a pressure wave in water for the non transformed part of the propagation medium. The transformed medium is composed of several concentric media with rigid elliptical-like perforations. Once again the elliptical-like perforations are designed in the $(r,\theta)$ coordinates and then mapped onto the $(x,y)$ coordinates using the conformal map introduced in equation (\ref{eq:TC}). The shape of the perforations is obtained using the genetic algorithm. The stopping criterion is a relative error of less than $5\%$, meaning $ \gamma (\alpha^H) < 0.05$ with input parameters identical to those introduced in figure \ref{fig2b}. The output parameters given by the GA for the different layers are summarized in table \ref{table2} and the geometry of the cloak is displayed in figure \ref{fig4a}. A qualitative numerical result is shown in figure \ref{fig4b} for a representative frequency $f = 13kHz$. When comparing tables \ref{table2} and \ref{table1} we notice that the rectangular lattice is less demanding with respect to the change in mass density and velocity. In fact, mass density for the laminar lattice goes from $47 kg/m^3$ to $21466 kg/m^3$, which seems unachievable in practice, whereas in the rectangular lattice case it goes from $37 kg/m^3$ to $717 kg/m^3$. A similar observation can be done for the velocity (from 831 m/s to 4654 m/s in laminar case, from 1495 m/s to 4734 m/s for rectangular case). When considering an experimental realization the rectangular lattice would then be more appropriate. 
\\
\FloatBarrier
\subsection{Quantitative measure of the efficiency of the cloaks}
\begin{figure}%
\includegraphics[width=\columnwidth]{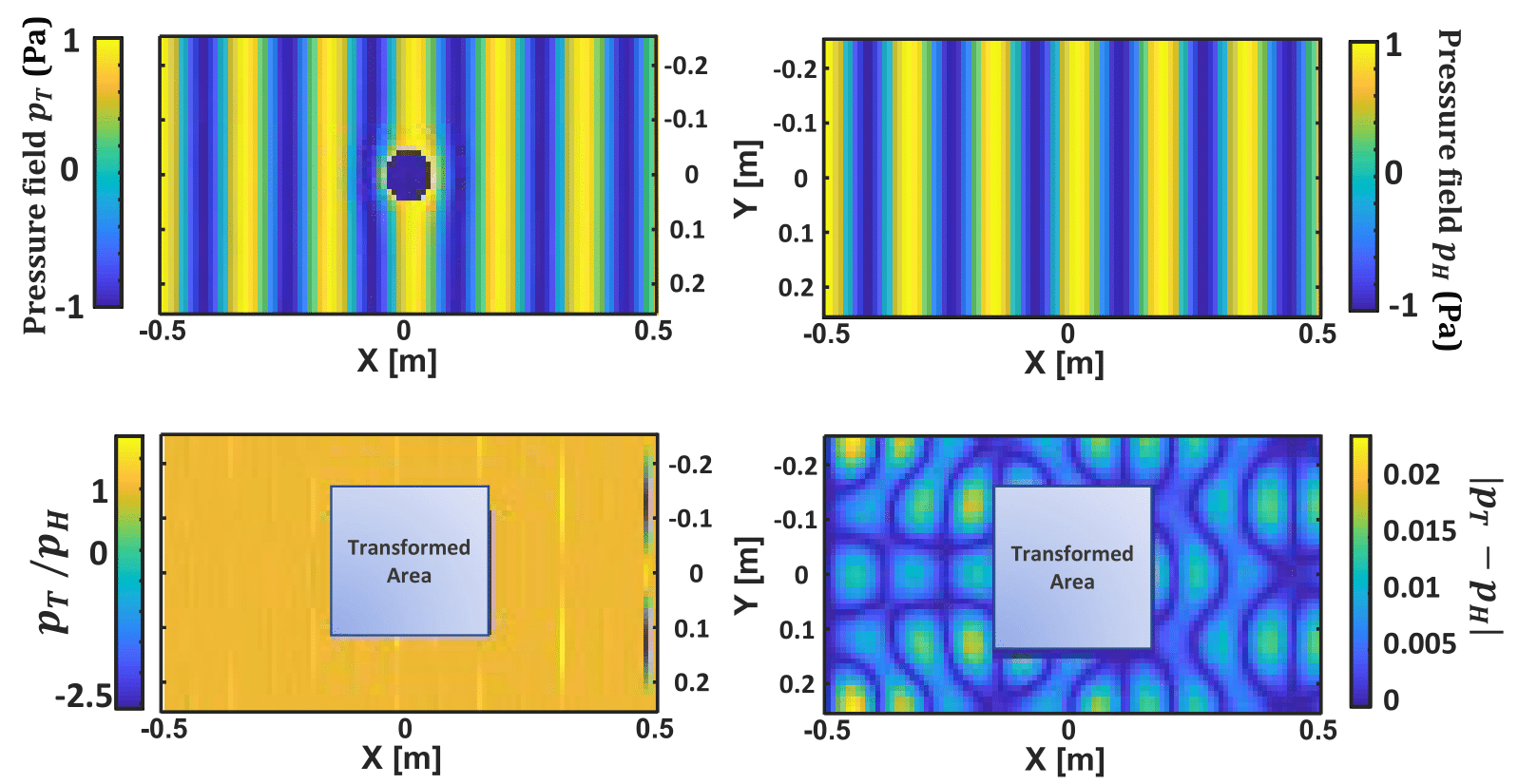}
\caption{Illustration of the procedures followed to measure the efficiency. First the pressure field in the transformed medium (upper left) is divided by same in the reference homogeneous medium (upper right). The result (bottom left) expresses the efficiency of the cloak at each points for a given frequency ($f = 8800$ Hz here). The closer the ratios to 1 the better the efficiency of the cloak. We then average the ratios computed at each point to obtain the overall efficiency at one frequency, and repeat this procedure for each frequency. We stress that due to resonance phenomena, meshing approximation or numerical approximation (for value very close to zero) artifacts can appear as we can see here on the rightmost side of the bottom left figure. The second criterion performs the difference between the transformed medium and the homogeneous medium, the result is shown on the bottom left of the figure. We can already observe that this criterion smooths out the artifacts and reveals some resonance pattern.}%
\label{fig:efficiencyProcedure}%
\end{figure}

\begin{figure*}
        \centering
        \begin{subfigure}[b]{0.475\textwidth}
            \centering
            \includegraphics[width=0.9\textwidth]{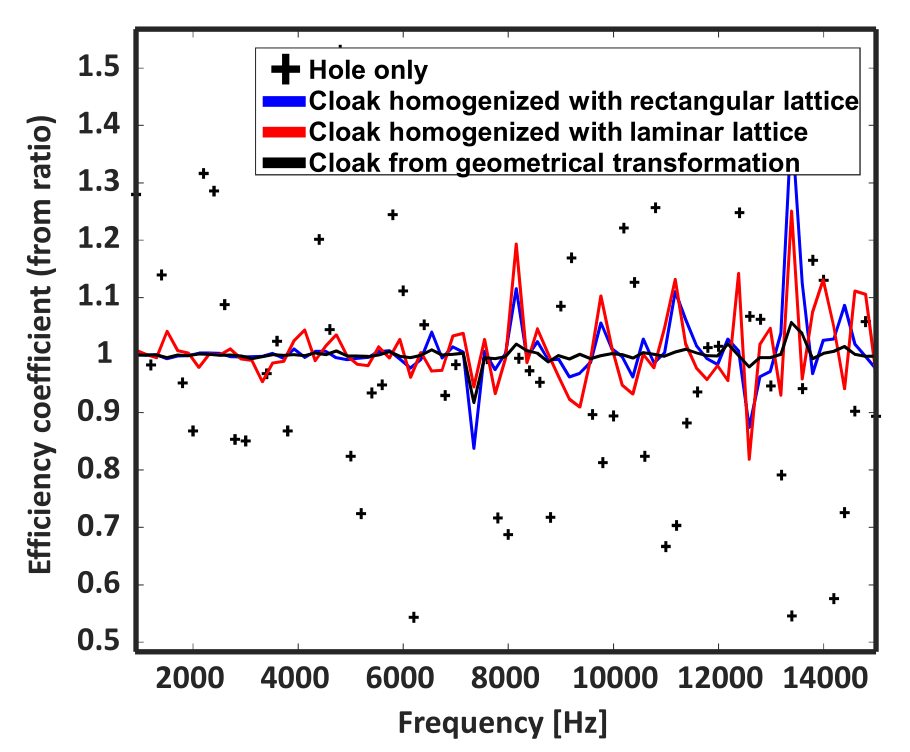}
            \caption{}  
            \label{fig:efficiencyRatio}
        \end{subfigure}
        \hfill
        \begin{subfigure}[b]{0.475\textwidth}  
            \centering 
            \includegraphics[width=0.9\textwidth]{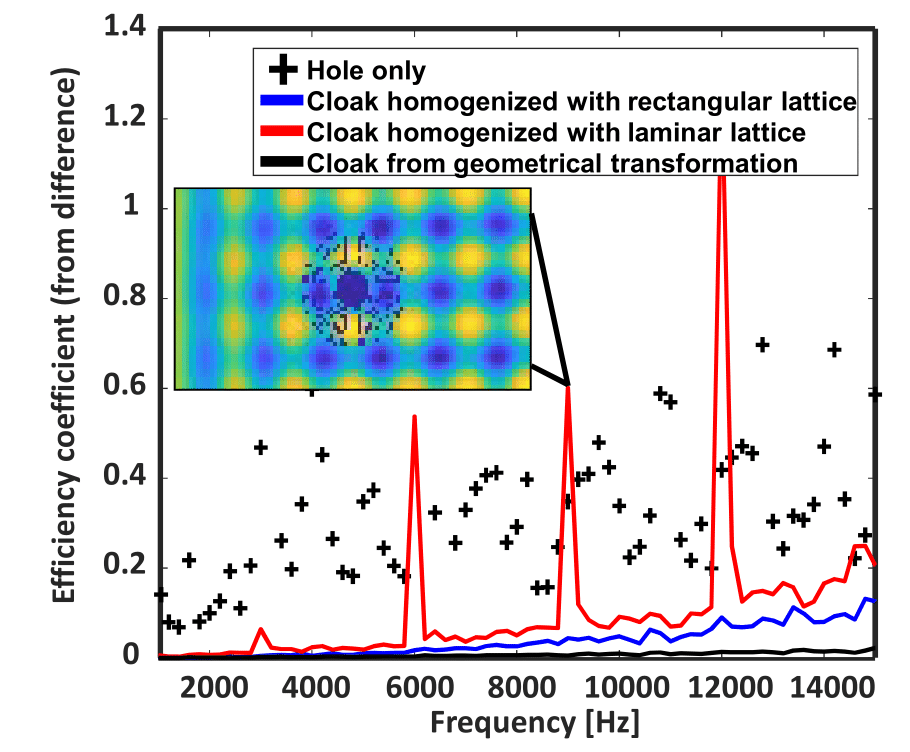}
            \caption{}   
            \label{fig:efficiencyDifference}
        \end{subfigure}
				\caption{The crosses show the value in the case without cloak (for comparison) as shown in figure \ref{figChoiceOfM}, the black line is the anisotropic cloak given by the transformation (figure \ref{fig1} (Lower Right)), the blue line the cloak homogenized with the rectangular lattice, the red line the cloak homogenized with the laminar lattice. a) Quantitative measure of the efficiency of each cloak for a large band of frequencies with a difference criterion defined by the ratio between the transformed case and the homogeneous case. A ratio close to one indicates a strong efficiency. b) Quantitative measure of the efficiency of each cloak for a large band of frequencies with a difference criterion defined by the difference between the transformed case and the homogeneous case. A ratio close to zero indicates a strong efficiency.}  
        \label{fig:efficiency}
\end{figure*}

In this section we describe the quantitative method that we choose to measure the efficiency of the cloaks depending on the frequency. To do so we compare the value of the total pressure field between the homogeneous case and the transformed cases under study at over 5000 points located outside the region of the cloak. At a given point, we perform a ratio between the two field values (see figure \ref{fig:efficiencyProcedure}) and we further average all the ratios. We repeat this procedure for each frequency. Obviously the closer the ratio to 1 the better the efficiency of the cloak. To complete the study we also consider an efficiency criterion defined by the difference between the two fields where the closer the difference to 0 the better the efficiency. \\
\FloatBarrier
As a benchmark we use the anisotropic inhomogeneous cloak presented in figure \ref{fig1} (Lower right) (ie the transformed medium without any approximation). We then compare the homogeneous case with the case introduced in figure \ref{figChoiceOfM} for $M=0$ in order to show the impact of the various cloak designs. We then compare the 1D laminar case (fig\ref{fig3}) and  the 2D rectangular lattice (fig\ref{fig4}) against these two benchmarks. All the results are summarized in figure \ref{fig:efficiencyRatio}. It should be noted that the black curve should in theory be equal to $1$ at every frequency. However, with the chosen criterion  singularities can occur due to vanishing denominator. Thus we give an alternative criterion in figure \ref{fig:efficiencyDifference} replacing a ratio by a difference. This has the advantage of removing the imperfections of the reference curve  but can none the less hide meaningful observations on cloak's efficiency.\\
\FloatBarrier
We observe from these quantitative results that the efficiency of all cloaks is extremely satisfactory on the low frequency range, but this efficiency decreases at higher frequencies for the homogenized cloaks (line blue and red). We can also note that the efficiency of the cloak made of a 2D rectangular lattice decreases faster than the 1D laminar case. This can be attributed to the fact that the elementary cells are smaller for the 1D laminar (~0.8mm) case than for the 2D rectangular lattice (5 mm), and moreover in the 2D case the Neumann (rigid) inclusions can introduce some multiple scattering. Furthermore if we look at the results given by the second criterion we observe the presence of resonant modes for the perforated medium that do not appear in the other cases. If the 2D rectangular lattice seems easier to reproduce in practice, one should also take into account the loss of efficiency due to the presence of resonant modes and the faster decrease in efficiency for higher frequency. 
\section{Additional micro-structures with exotic elementary cell}
\begin{figure}%
\includegraphics[width=\columnwidth]{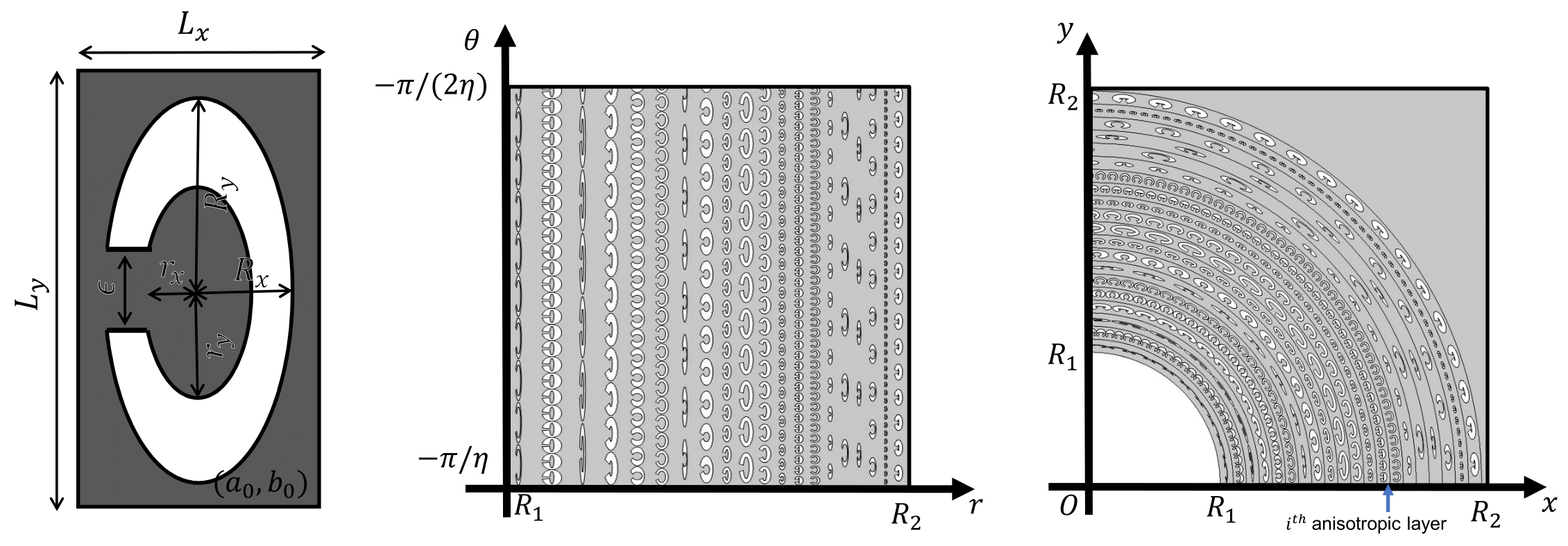}
\caption{Left panel: the elementary cell, defined by 9 parameters. Middle panel: the microstructure before the conformal mapping, $\eta$ is defined in equation (\ref{eq:TC}). Right panel: the microstructure after the conformal mapping.}%
\label{fig:croissantMS}%
\end{figure}

\begin{figure*}
        \centering
        \begin{subfigure}[b]{0.9\textwidth}
            \centering
            \includegraphics[width=\textwidth]{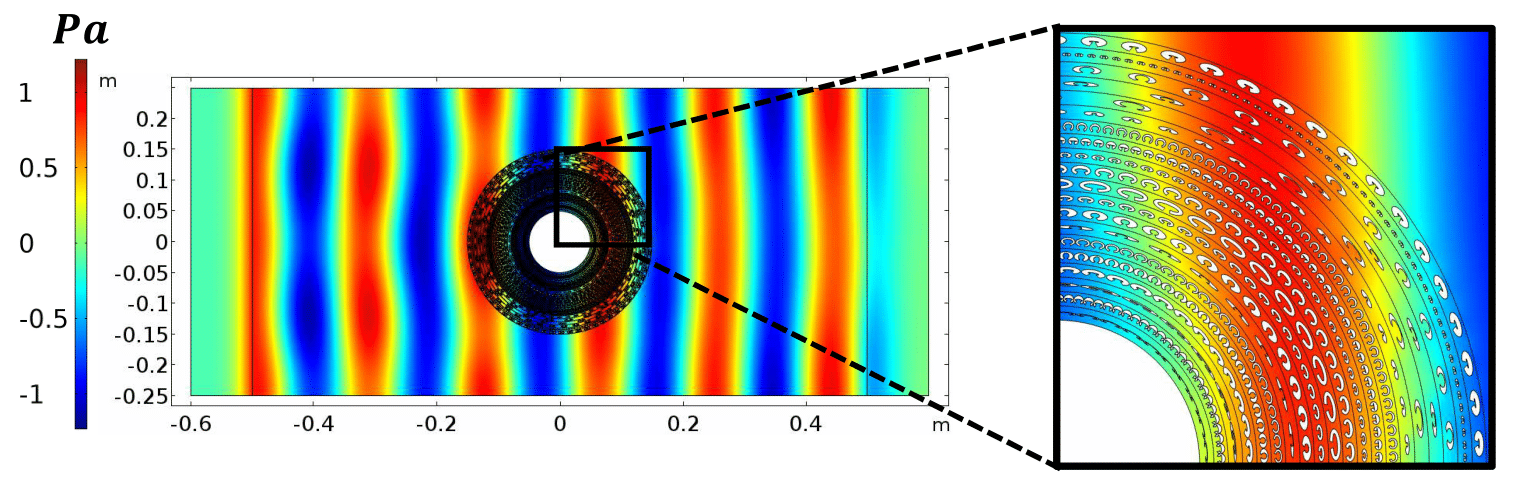}
            \caption{}  
            \label{fig:croissantResultsa}
        \end{subfigure}
%        \baselineskip
        \begin{subfigure}[b]{0.45\textwidth}  
            \centering 
            \includegraphics[width=\textwidth]{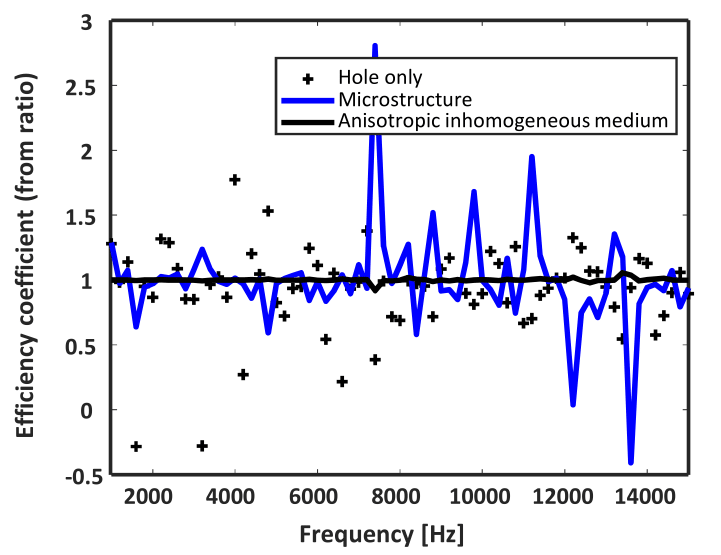}
            \caption{}   
            \label{fig:croissantResultsb}
        \end{subfigure}
				\begin{subfigure}[b]{0.45\textwidth}  
            \centering 
            \includegraphics[width=\textwidth]{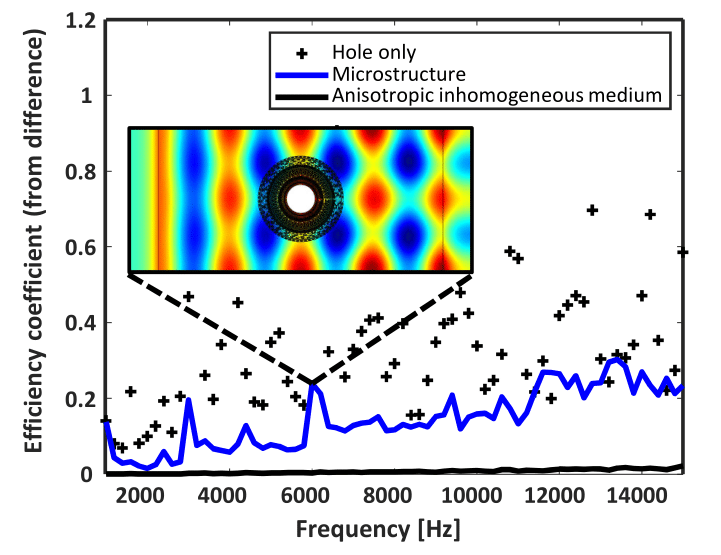}
            \caption{}   
            \label{fig:croissantResultsc}
        \end{subfigure}
				\caption{a) Numerical result at a given frequency ($f = 8000$ Hz) . b) Quantitative efficiency obtained with the ratio criterion. c) Quantitative efficiency obtained with the difference criterion. We can observe the drop in the efficiency attributed to the resonance phenomena.}  
        \label{fig:croissantResults}
\end{figure*}

\begin{figure}%
\includegraphics[width=\columnwidth]{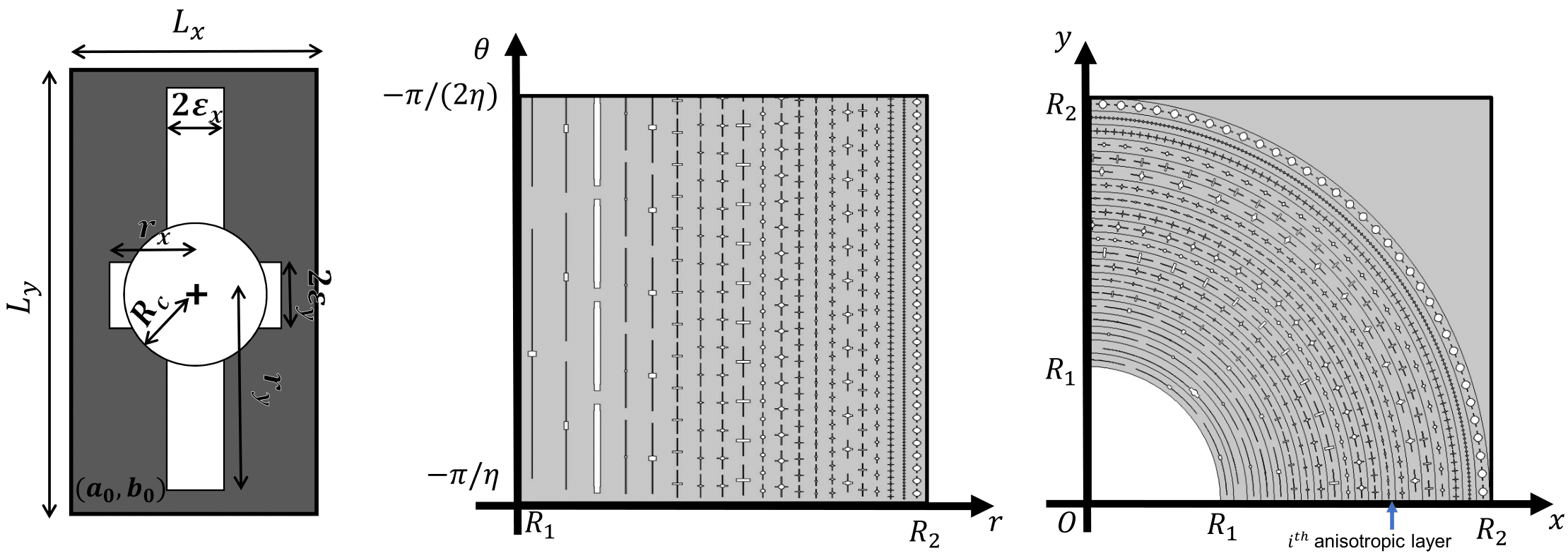}
\caption{Left panel: the elementary cell, defined by 9 parameters. Middle panel: the microstructure before the conformal mapping. Right panel: the microstructure after the conformal mapping, see Eq. (\ref{eq:TC}), where the parameter $\eta$ is defined.}%
\label{fig:celticCrossMS}%
\end{figure}

\begin{figure*}
        \centering
        \begin{subfigure}[b]{0.9\textwidth}
            \centering
            \includegraphics[width=\textwidth]{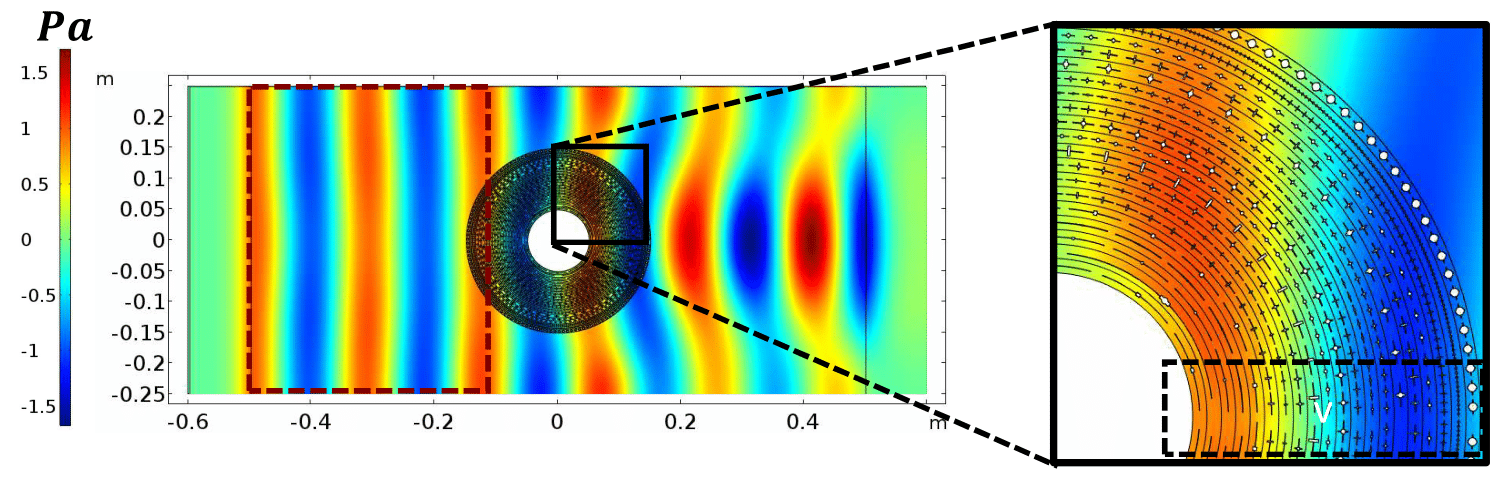}
            \caption{}  
            \label{fig:celticCrossResulta}
        \end{subfigure}
%        \baselineskip
        \begin{subfigure}[b]{0.45\textwidth}  
            \centering 
            \includegraphics[width=\textwidth]{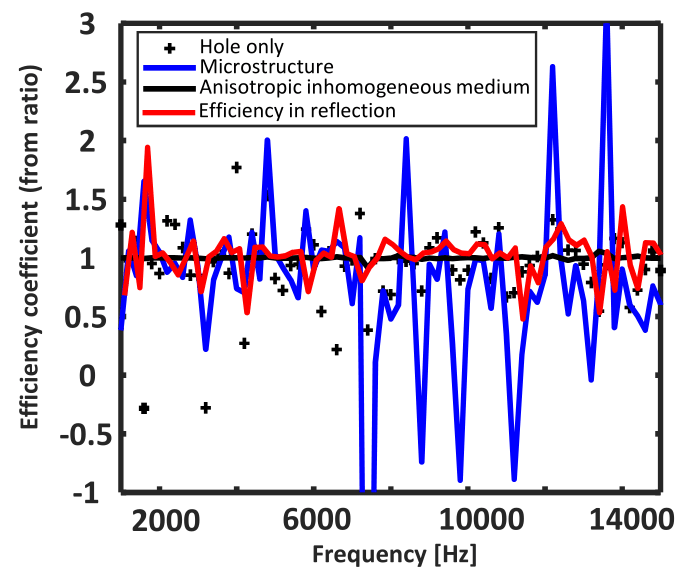}
            \caption{}   
            \label{fig:celticCrossResultb}
        \end{subfigure}
				\begin{subfigure}[b]{0.45\textwidth}  
            \centering 
            \includegraphics[width=\textwidth]{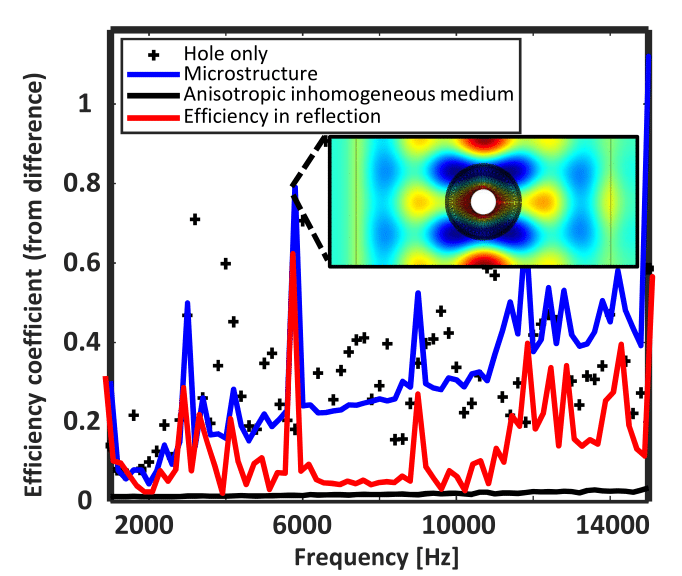}
            \caption{}   
            \label{fig:celticCrossResultc}
        \end{subfigure}
				\caption{a) Numerical result at a given frequency ($f = 8000$ Hz). Looking at the zoom on the microstructure we observe some design imperfections as the perimeters of the layers are not a multiple of the size of the associated elementary cell. The left part of the design being not influenced by the design imperfections the cloak does not produce any reflections. The right part introduces a distortion in the wavefront as it contains the imperfections. b) Quantitative efficiency obtained with the ratio criterion. c) Quantitative efficiency obtained with the difference criterion. We observe the drop in the efficiency attributed to the resonance phenomena and the imperfections in the microstructure.}  
        \label{fig:celticCrossResult}
\end{figure*}

To challenge the efficiency of the method, we further design several microstructures with more exotic elementary cells. The first one we want to consider is a microstructure consisting of split ring resonators as in \cite{Pendry1999}, which seems to be both appropriate to obtain strong anisotropy, and also is reminiscent of the design of the first electromagnetic cloak \cite{Schurig2006}. The elementary cell of this structure and its geometric parameters are introduced in figure \ref{fig:croissantMS}. We note that this elementary cell is defined by 9 parameters, whereas the elementary cell with an elliptic perforation was defined by 6 parameters only (and the elementary cell for homogeneous layers required only 4 parameters). Thus, it seems to us that the genetic algorithm is absolutely necessary in the present case. The resulting microstructure before and after the conformal map is shown on figure \ref{fig:croissantMS} and all the geometrical and physical parameters are summarized in table \ref{table:croissant}. \\
We then perform some numerical simulation on this microstructure. The result for one frequency and the efficiency curves for both criteria (ratio and difference between pressure fields in homogeneous and microstructured media, respectively) are shown on figure \ref{fig:croissantResults}. We observe  from the reading of figure \ref{fig:croissantResults} and figure \ref{fig:efficiency} that the overall cloak's efficiency is not as good as for the previous microstructures which we studied. This can be attributed to the form of the inclusions that are more prone to collective resonant behavior. As we have seen before for the elliptic microstructure, resonance phenomena decrease the efficiency of the cloak. In actuality, upon inspection of figure \ref{fig:croissantResultsb}, it can be seen that the cloak only works properly at frequencies lower than 7000 Hz (and similarly for figure \ref{fig:efficiencyRatio}): clearly, the homogenization theory breaks down at frequencies above 7000 Hz. The worsening of the cloak efficiency in the split ring case can be easily explained by the fact that the homogenization method which we consider here does not take into account local resonance phenomena, thus the elementary cells mimic properly the transformed medium at all frequencies except for the resonant ones, as long as the ratio of unit cell size to wave wavelength remains small enough. \\
We continue to challenge our method with a new design of elementary cell defined again by 9 parameters. We choose this design as it seems unlikely an effective medium approach could be applied to the Celtic cross within the elementary cell (shown on figure \ref{fig:celticCrossMS}): we call it a Celtic cross elementary cell. Once again the elementary cell is defined by 9 parameters. We display the microstructured design before and after the geometric transformation on figure \ref{fig:celticCrossMS}. All the parameters of this microstructure are summarized in table \ref{table:celticCross}. \\
We again perform numerical simulations on this microstructure and determine the efficiency curves depending on the frequency. We show the results on figure \ref{fig:celticCrossResult}. We observe that the qualitative and quantitative results decline in comparison with our previous designs. On the positive side, we observe that the cloak does not produce any reflections (see leftmost side of figure \ref{fig:celticCrossResulta} and red curves in figures \ref{fig:celticCrossResultb},\ref{fig:celticCrossResultc}). On the negative side, the wavefront is more distorted after the cloak than it used to be in the previous designs, so the overall cloak's efficiency worsens. The poor cloaking efficiency can be explained by the fact that the perimeters of the various circular homogeneous anisotropic layers are not an integer multiple of the associated elementary cells, as can be observed in the inset showing a magnified view in figure \ref{fig:celticCrossResulta}. The salient consequence is a default in the microstructure. This default was also present in the previous design but had less influence. However by comparing the qualitative result of figure \ref{fig:celticCrossResulta} and the qualitative result in the case with no cloak (see figure \ref{figChoiceOfM}, upper right) we still observe a clear improvement of the invisibility due to the cloak.
\section{Concluding remarks}
In this paper we describe a general method to mimic complex anisotropic inhomogeneous media obtained through geometrical transformation using classical homogenization results combined with a genetic algorithm in the case of the Helmholtz equation. First the complex medium is created mathematically using the form invariance property of the Helmholtz equation, then this theoretical material is approximated by performing what we call inverse homogenization as a reference to \cite{Cherkaev2001}. This method can be easily adapted to several areas of wave physics such as acoustics or electromagnetism, where governing equations are form invariant (with the notable constraint that one cannot use the conformal mapping for 3D cloaks, and thus the mapping in Eq. (\ref{eq:TC}), would need to be replaced by another, more complicated, one \cite{Francesco1997}. Indeed, in dimension 3, the wanted conformal map should be a compound transform deduced from a homothetic transformation, an isometry and a special conformal transform (the latter being the composition of a reflection and the inversion of a sphere, such as a Moebius transform). \\
Our method would require further generalization to tackle elasticity cases, for which the form invariance of the governing equations is less obvious \cite{Norris2011,Pomot2019, Colquitt2014}. Considering the versatility and the room for further improvement of our method (e.g. multiphase/multiscale media and 3D designs), we believe that our approach will pave the way to further numerical studies as well as foster efforts into experimental realizations.

\bibliography{biblio}

\appendix
\section{Tables of parameters for the various microstructures}
\begin{table}
\centering
\caption{Numerical values of the mass density and velocity used in the design of a cloak made of 5 homogeneous anisotropic layers, each of them made of 3 alternations of two homogeneous isotropic layers.}
\begin{tabular}{|c|c|c|c|c|}
  \hline
  $i^{th}$ layer & $\rho^1_i$ [kg/m$^3$] & $c^1_i$ [m/s] & $\rho^2_i$ [kg/m$^3$] & $c^2_i$ [m/s] \\
  \hline
  1 ($r = R_1$) & 6389 & 559 & 156 & 3575 \\
  2 & 2942 & 824 & 340 & 2426 \\
	3 & 2215 & 950 & 451 & 2105 \\
	4 & 1891 & 1028 & 528 & 1945\\
	5 ($r = R_2$)& 1706 & 1083 & 586 & 1847 \\
  \hline
	\label{table1forM5}
\end{tabular}
\end{table}

\begin{table}
\centering
\caption{Numerical values of the mass density and velocity used in the design of the cloak introduced in figure \ref{fig3}.}
\begin{tabular}{|c|c|c|c|c|}
  \hline
  $i^{th}$ layer & $\rho^1_i$ [kg/m$^3$] & $c^1_i$ [m/s] & $\rho^2_i$ [kg/m$^3$] & $c^2_i$ [m/s] \\
  \hline
  1 ($r = R_1$) & 21466 & 2261 & 47 & 4654 \\
  2 & 8080 & 2145 & 124 & 2881 \\
	3 & 5369 & 2034 & 186 & 2371 \\
	4 & 4193 & 1930 & 238 & 2117\\
	5 & 3531 & 1831 & 283 & 1964 \\
	6 & 3104 & 1737 & 322 & 1864 \\
	7 & 2805 & 1648 & 356 & 1797 \\
	8 & 2582 & 1563 & 387 & 1752 \\
	9 & 2411 & 1483 & 415 & 1724 \\
	10 & 2274 & 1407 & 440 & 1710 \\
	11 & 2162 & 1335 & 462 & 1709 \\
	12 & 2069 & 1266 & 483 & 1721 \\
	13 & 1990 & 1201 & 502 & 1747 \\
	14 & 1922 & 1140 & 520 & 1791 \\
	15 & 1863 & 1081 & 537 & 1856 \\
	16 & 1811 & 1026 & 552 & 1952 \\
	17 & 1766 & 973 & 566 & 2096 \\
	18 & 1725 & 923 & 580 & 2323 \\
	19 & 1689 & 876 & 592 & 2722 \\
	20 ($r = R_2$) & 831 & 1099 & 604 & 3639\\
  \hline
	\label{table1}
\end{tabular}
\end{table}

\begin{table}
\centering
\caption{ Numerical values of the mass density, velocity and parameters of the elementary cells used in the design of the cloak introduced in figure \ref{fig4}. The comparison with table \ref{table1} show that the rectangular lattice would be experimentally less demanding than the laminar lattice.}
\begin{tabular}{|c|c|c|c|c|c|c|}
  \hline
  $i^{th}$ layer & $\rho_i$ [kg/m$^3$] & $c_i$ [m/s] & $L_x^i$ [mm] & $L_y^i [mm]$ & $r_x^i$ [mm] & $r_y^i$ [mm] \\
  \hline
  1 ($r = R_1$) & 37 & 4734 & 8.7 & 76.9 & 2.9 & 38\\
  2 & 182 & 2623 & 7.9 & 5.9 & 1.5 & 2.9\\
	3 & 338 & 2156 & 7.3 & 103.7 & 0.7 & 42.8\\
	4 & 422 & 1919 & 6.7 & 8.3 & 0.6 & 3.9\\
	5 & 376 & 1912 & 6.3 & 25.9 & 1.1 & 10.7\\
	6 & 205 & 1944 & 5.9 & 6.6 & 2.2 & 3.1\\
	7 & 298 & 1782 & 5.5 & 7.3 & 1.7 & 3.4\\
	8 & 414 & 1701 & 5.2 & 6.8 & 1.2 & 3\\
	9 & 597 & 1588 & 4.9 & 7 & 0.6 & 2.7\\
	10 & 548 & 1604 & 4.7 & 4.3 & 0.7 & 1.8\\
	11 & 473 & 1592 & 4.4 & 3.3 & 1.1 & 1.4\\
	12 & 625 & 1544 & 4.2 & 6.4 & 0.6 & 2.2\\
	13 & 306 & 1686 & 4 & 5.9 & 1.5 & 2.6\\
	14 & 608 & 1519 & 3.9 & 1.2 & 0.6 & 0.5\\
	15 & 548 & 1535 & 3.7 & 3.8 & 0.9 & 1.4\\
	16 & 717 & 1495 & 3.6 & 3.3 & 0.3 & 1.1\\
	17 & 260 & 1736 & 3.4 & 2.7 & 1.4 & 1.2\\
	18 & 468 & 1574 & 3.3 & 2.2 & 0.9 & 0.8\\
	19 & 480 & 1564 & 3.2 & 3.2 & 0.9 & 1.2\\
	20 ($r = R_2$) & 446 & 1605 & 3.1 & 2.7 & 0.9 & 1.1\\
  \hline
	\label{table2}
\end{tabular}
\end{table}

\begin{table}
\centering
\caption{ Numerical values of the mass density, velocity and parameters of the elementary cells used in the design of the cloak introduced in figure \ref{fig:croissantMS}}
\begin{tabular}{|c|c|c|c|c|c|c|c|c|c|}
  \hline
  $i^{th}$ layer & $\rho_i$ [kg/m$^3$] & $c_i$ [m/s] & $L_x^i$ [mm] & $L_y^i$ [mm] & $R_x^i$ [mm] & $R_y^i$ [mm] & $\epsilon^i$ & $r_x^i$ [mm]& $r_y^i$ [mm] \\
  \hline
  %1 ($r = R_1$) & 102 & 4263 & 8.7 & 14.5 & 0.9 & 0.3 &  &  \\
	2 & 137 & 3664 & 7.9 & 5.6 & 2.4 & 2.8 & 2.9 & 0.5 & 2.1 \\
  3 & 320 & 2458 & 7.3 & 21.8 & 0.7 & 10.5 & 3.6 & 0.3 & 6.2 \\
  4 & 313 & 2485 & 6.7 & 11.9 & 1.7 & 5.7 & 4.5 & 0.3 & 2.6 \\
	5 & 292 & 2451 & 6.3 & 6.8 & 1.8 & 3.2 & 2.1 & 1.2 & 1.3 \\
	6 & 349 & 2193 & 5.9 & 6.1 & 1.6 & 2.9 & 2 & 0.9 & 2 \\
	7 & 572 & 1832 & 5.5 & 18.1 & 0.6 & 6.9 & 5.4 & 0.5 & 2.4 \\
	8 & 380 & 2214 & 5.2 & 11.2 & 1.6 & 4.6 & 2 & 0.8 & 1.6 \\
	9 & 501 & 1951 & 4.9 & 7.2 & 1.1 & 2.8 & 1.4 & 0.3 & 1.7\\
	10 & 321 & 2308 & 4.7 & 12.4 & 1.7 & 5.2 & 2.2 & 0.7 & 3.7 \\
	11 & 409 & 2104 & 4.4 & 7.3 & 1.5 & 3 & 2.8 & 0.5 & 2 \\
	12 & 571 & 1828 & 4.2 & 4.9 & 0.8 & 1.9 & 0.8 & 0.3 & 0.9 \\
	13 & 490 & 2002 & 4.0 & 3.6 & 1.1 & 1.4 & 0.9 & 0.1 & 1.1 \\
	14 & 468 & 1918 & 3.9 & 3.6 & 1.2 & 1.4 & 1.3 & 0.6 & 1 \\
	15 & 722 & 1655 & 3.7 & 9.9 & 0.5 & 2.6 & 1.7 & 0.2 & 0.6 \\
	16 & 699 & 1621 & 3.6 & 22.5 & 0.9 & 4.7 & 3.7 & 0.7 & 3 \\
	17 & 750 & 1611 & 3.4 & 21.1 & 0.5 & 4.2 & 1.8 & 0.4 & 1.8 \\
	18 & 655 & 1675 & 3.3 & 12.9 & 0.9 & 3.2 & 2.8 & 0.7 & 1.7 \\
	19 & 761 & 1598 & 3.2 & 2.8 & 0.3 & 0.9 & 0.5 & 0.2 & 0.4 \\
	20 ($r = R_2$) &  511 & 1965 & 3.1 & 8.7 & 0.1 & 2.8 & 1.2 & 0.1 & 1.1\\
  \hline
	\label{table:croissant}
\end{tabular}
\end{table}

\begin{table}
\centering
\caption{ Numerical values of the mass density, velocity and parameters of the elementary cells used in the design of the cloak introduced in figure \ref{fig:celticCrossMS}}
\begin{tabular}{|c|c|c|c|c|c|c|c|c|c|}
  \hline
  $i^{th}$ layer & $\rho_i$ [kg/m$^3$] & $c_i$ [m/s] & $L_x^i$ [mm] & $L_y^i$ [mm] & $r_x^i$ [mm] & $r_y^i$ [mm] &  $\epsilon_x^i$ [mm]& $\epsilon_y^i$ [mm] & $R_c$ \\
  \hline
  1 ($r = R_1$) & 107 & 3990 & 8.7 & 56.8 & 0.9 & 24.6 & 0.08 & 1.3 & 1 \\
	2 & 248 & 2537 & 6.3 & 142.7 & 0.9 & 61.3 & 0.07 & 3.5 & 1.8  \\
	3 & 379 & 2231 & 7.3 & 37.3 & 0.3 & 16.1 & 0.06 & 0.5 & 0.7   \\
	4 & 449 & 2051 & 6.7 & 28.2 & 0.3 & 12.1 & 0.12 & 0.3 & 0.3  \\
	5 & 488 & 1814 & 6.3 & 27.2 & 0.9 & 11.6 & 0.08 & 0.7 & 0.8  \\
	6 & 520 & 1664 & 5.9 & 10.5 & 1.2 & 4.5 & 0.08 & 0.3 & 0.3   \\
	7 & 621 & 1633 & 5.5 & 9.1 & 0.7 & 3.8 & 0.05 & 0.1 & 0.4   \\
	8 & 577 & 1549 & 5.2 & 9.1 & 1.3 & 3.6 & 0.1 & 0.2 & 0.5   \\
	9 & 536 & 1480 & 4.9 & 13 & 1.7 & 5 & 0.09 & 0.4 &  0.5 \\
	10 & 675 & 1586 & 4.7 & 5.9 & 0.6 & 2.2 & 0.05 & 0.1 & 0.6   \\
	11 & 563 & 1428 & 4.4 & 8.5 & 1.6 & 3.3 & 0.12 & 0.1 & 0.7  \\
	12 & 666 & 1455 & 4.2 & 5.4 & 1 & 2.1 & 0.09 & 0.03 & 0.3\\
	13 & 788 & 1538 & 4 & 4.1 & 0.2 & 1.5 & 0.07 & 0.02 & 0.3   \\
	14 & 733 & 1468 & 3.9 & 4.5 & 0.7 & 1.6 & 0.05 & 0.1 & 0.3   \\
	15 & 685 & 1459 & 3.7 & 11.1 & 1.2 & 3 & 0.06 & 0.3 & 0.7  \\
	16 & 690 & 1442 & 3.6 & 6.6 & 1 & 2.1 & 0.09 & 0.2 & 0.3  \\
	17 & 826 & 1437 & 3.4 & 5.4 & 0.6 & 1.5 & 0.04 & 0.04 & 0.4   \\
	18 & 678 & 1458 & 3.3 & 2.1 & 0.7 & 0.9 & 0.06 & 0.02 & 0.2   \\
	19 & 786 & 1490 & 3.2 & 1 & 0.3 & 0.4 & 0.02 & 0.05 & 0.2   \\
	20 & 560 & 1494 & 3.1 & 4.2 & 1 & 1.5 & 0.08 & 0.2 & 0.9 \\
  \hline
	\label{table:celticCross}
\end{tabular}
\end{table}

\end{document}